\documentstyle[prc,aps,epsfig,preprint]{revtex}
\begin{document}
\draft
\newcommand{\bsigma}{\mbox{\boldmath $\sigma$}}
\newcommand{\btau}{\mbox{\boldmath $\tau$}}
 
\title{Single particle properties of $^{16}$O and $^{40}$Ca}

\author{
A.\ Fabrocini
}
\address{INFN, Sezione di Pisa, I-56100 Pisa, Italy and}
\address{Department of Physics, University of Pisa, I-56100 Pisa, Italy}
\date{\today}
\maketitle
\begin{abstract}
We discuss some single particle properties 
 of $^{16}$O and $^{40}$Ca using correlated basis function theory 
and Fermi hypernetted chain equations 
with central and tensor correlations. In particular, we 
concentrate on one body density matrix, 
momentum distribution, natural orbits and quasi hole states.
The correlations are variationally generated by a 
 realistic  hamiltonian containing the Argonne $v'_{8}$ two--nucleon 
and Urbana IX  three--nucleon interactions. 
The correlated  momentum distributions show 
 the well known enhancement at large momenta, with  
a relative importance of the different correlations 
(Jastrow and tensor) similar to that in nuclear matter. 
The natural orbits and their occupation numbers are obtained by  
diagonalization of the density matrix.  The correlated first natural 
orbits occupations are depleted 
 by more than 10$\%$, whereas the first following ones are 
 occupied by a few percent. The spectroscopic factors of the valence states 
are lowered by $\sim 8-12\%$  with respect to unity by central and 
tensor correlations, confirming that short range correlations alone 
are not able to explain the values extracted from $(e,e'p)$ experiments. 
\end{abstract}

\section{Introduction}

The behavior of the single nucleon in the medium bears  
clear signatures of the nucleon--nucleon (NN) correlations  induced by the 
strong nuclear interaction. 
The most celebrated one is the presence of high 
momentum components in the momentum distribution (MD), which may 
be explained only in terms of  short range correlations (SRC) and 
cannot be justified by any independent particle model (IPM)\cite{zab78,ant88}. 
In addition, both short-- and long--range correlations (LRC) are now 
widely thought to be the origin of the large  reduction of the spectroscopic strengths 
of the hole states\cite{ste91}. 

In general, the investigation of SRC is one of the main topics in the physics of 
strongly interacting 
systems, from liquid Helium to nuclear matter. The translational invariance of homogeneous 
media has made possible a quantitative assesment of their importance in these contexts. 
However, the advances in realistic many--body theories have only recently  allowed 
to address the same issue in inhomogeneous objects, as atomic nuclei, with a comparable 
accuracy.  To this aim, correlated basis functions (CBF) theory has emerged as an effective
 tool to tackle the SRC problem in nuclear systems \cite{wir88} when realistic 
hamiltonians are employed. When used in conjuction with the Fermi hypernetted chain (FHNC) 
integral equations\cite{co92,ari96}, CBF has attained in $^{16}$O and $^{40}$Ca  
the same accuracy as in the best variational studies of nuclear matter\cite{fab98,fab00}. 

In order to dissentagle the SRC effects, one has to look at effects that go beyond any 
possible IPM description of the nucleus. SRC are mainly produced 
by the repulsive core of the NN interaction, whereas important intermediate 
and long range contributions (essentially of the tensor type) come from the exchange of 
one  or more pions. This part  of the nuclear interaction is the responsible of 
the nuclear binding\cite{fab98}. Direct evidence of the SRC are not easy to  
experimentally single out and only in these last few years, thanks to the advances in the
experimental techniques, there have been consistent 
efforts aimed to their identification. For instance, $(e,e'p)$ data in the quasi-elastic 
region need a large reduction of the IPM hole strength to be reproduced \cite{dew90}. 
Moreover, charge density distributions obtained by elastic electron
scattering experiments are smaller than 
those predicted by the IPM \cite{cav82} in the nuclear interior. 
These facts can be be explained by assuming occupation probabilities  of the single particle 
levels different from the IPM ones\cite{pap86}.

The basic quantity to investigate in order to verify the hypothesis of partial occupation
probability is the one body density matrix (OBDM), 
$\rho({\bf r}_1,{\bf r}_{1'})$, defined as:
\begin{equation}
\rho({\bf r}_1,{\bf r}_{1'})=
\langle \Psi_0(A) \vert a^\dagger({\bf r}_1) a({\bf r}_{1'})
       \vert \Psi_0(A) \rangle
\, ,
\label{OBDM_0}
\end{equation}
where $\Psi_0(A)$ is the ground state A--body wave function and 
$a^\dagger(a)({\bf r}_1)$ is the creation (annihilation) operator of a nucleon  
at the position ${\bf r}_1$. The Fourier transform of the OBDM gives 
the momentum distribution (MD), $n(k)$, sometimes  
used in plane wave impulse approximation  to study 
inclusive and exclusive reactions. 
 The natural orbits (NO)\cite{mau91}, with their occupation numbers 
($n_\alpha$),  are defined as  the basis where the OBDM is diagonal. 
In the IPM, the nuclear ground state is described by 
a Slater determinant of fully occupied single particle (SP) wave functions 
below the Fermi surface, $\alpha_F$. So, the NO and the SP w.f. 
coincide, with $n_{\alpha\leq \alpha_F}$=1 and $n_{\alpha > \alpha_F}$=0. 
 Deviations from this situation are a measure of the correlations, 
since they deplete the populated orbitals and allow higher NO to get 
$n_{\alpha > \alpha_F}\neq$0. 

Other important quantities are the quasihole (QH) wave functions, $\psi_h({\bf r})$, 
defined as the overlaps  between $\Psi_0$ and the hole states, 
$\Psi_h$,  obtained by removing a nucleon from the position ${\bf r}$.
 From $(e,e'p)$ experiments it is possible to obtain an accurate determination
of the QH overlap functions\cite{kel96}, whose  normalizations give the
spectroscopic factors, $S_h$. 
Typical values of $S_h$, as extracted from the experiments,  are
$S_h\sim 0.6-0.7$\cite{lap93}, and the deviations from unity (their IPM value) 
come in account of various effects,  from center of mass to correlation corrections.

In CBF theory the description of the correlations starts with an 
A--body correlated wave function
\begin{equation}
\label{ansatz}
\Psi_0(1,2...A)= {\cal S}(\prod_{i<j}F_{ij}) \Phi_0(1,2...A)
\, , 
\label{Psi_0}
\end{equation}
corresponding to a 
symmetrized product of non commuting two-body correlation operators, $F_{ij}$, 
 acting on the  mean field wave function, $\Phi_0(1,2...A)$, given by a Slater 
determinant of  single particle wave functions, $\phi_\alpha(i)$. 
A realistic choice of $F_{ij}$ is:
\begin{equation}
F_{ij}=\sum_{p=1,6}f^p(r_{ij})O^p_{ij}
\, , 
\label{corr6}
\end{equation}
where
\begin{equation}
\label{oper6}
O^{p=1,8}_{ij}=
\left[ 1, \bsigma_i \cdot \bsigma_j, S_{ij}, 
({\bf L} \cdot {\bf S})_{ij} \right]\otimes
\left[ 1, \btau_i \cdot \btau_j \right] 
\end{equation}
and $S_{ij}=(3\,{\hat  {\bf r} }_{ij} \cdot \bsigma_i  \,
{\hat{\bf r}}_{ij} \cdot 
\bsigma_j -  \bsigma_i \cdot \bsigma_j)$ is the tensor operator. 

The variational principle provides a natural recipe to determine the correlation functions, 
$f^p(r)$, and the single particle wave functions by minimizing the 
ground state energy. In our calculation we adopt a non relativistic nuclear 
hamiltonian of the form:
\begin{equation}
H={{-\hbar ^2}\over2\,m}\sum_i\nabla_i^2+\sum_{i<j}v_{ij}
+\sum_{i<j<k}v_{ijk} 
\, , 
\label{hamilt}
\end{equation}
where we have used the $v'_8$ reduction of the Argonne $v_{18}$ \cite{wir95} potential 
and the  Urbana IX\cite{pud97} three--nucleon interaction.

\section{One body density matrix, momentum distribution and natural orbits} 

The one body density matrix (\ref{OBDM_0}) 
of doubly closed shell nuclei in ($ls$) coupling can be evaluated in 
FHNC\cite{fab01}, starting from 
\begin{equation}
\rho({\bf r}_1,{\bf r}_{1'})={A\over {\cal N}}
\,\int d^3r_2... \int d^3r_A\,
\Psi^\dagger_0(1,2,..A)\Psi_0(1',2,..A)\, ,
\label{OBDM}
\end{equation}
where ${\cal N}=\int d^3r_1... \int d^3r_A \vert \Psi_0 \vert ^2$ and $A$ is the 
mass number. Once the OBDM is known, its diagonal part 
provides the one body density, $\rho_1({\bf r}_1)$, and 
 the momentum distribution may be computed by 
\begin{equation}
n(k)={1 \over A}\, \int d^3r_1\, \int d^3r_{1'}\,  
\rho({\bf r}_1,{\bf r}_{1'})\, 
e^{i{\bf k} \cdot ({\bf r}_1-{\bf r}_{1'})}\, .
\label{MD}
\end{equation}
The independent particle model expression for the OBDM is well known and reads as:
\begin{equation}
\rho_{IPM}({\bf r}_1,{\bf r}_{1'})=
\sum_\alpha \phi_\alpha^\dagger(1) \phi_\alpha(1')\, .
\label{OBDM_IPM}
\end{equation}

In the case of Jastrow correlated wave functions (when only the first 
component of the correlation (\ref{corr6}) is retained), 
the density matrix may be expanded in powers 
of the {\sl dynamical correlations}, $h(r)=[f^1(r)]^2-1$ 
and $\omega(r)=f^1(r)-1$, and of the {\sl statistical 
correlations}.  The expansion generates  cluster terms classified according to the 
number of particles and to the number of the correlation lines. 
The FHNC equations allow for summing cluster terms at all orders. 
Details of the finite systems FHNC theory may be found in Ref.\cite{co92,co94}.  
  The FHNC equations for the more general $f_6$ correlation 
of (\ref{corr6}) were derived in Ref.\cite{fab98,fab01} for the one and two 
body densities and for the OBDM. 

The single particle wave functions needed to build the Slater determinant 
 $\Phi_0(1,2...A)$ have been obtained by solving 
the single particle Schr\"odinger equation with a Woods--Saxon potential, 
\begin{equation}
V_{WS}(r)= {V_0 \over {1+\exp{[(r-R_0)/a_0]}}}\, ,
\label{WS}
\end{equation}
whose parameters have been fixed to reproduce at 
best the nuclei empirical densities. This choice, while spoiling the 
binding energies by only a few percent with respect to a full minimization, 
provides an accurate description of the nuclear densities\cite{fab00}.

\begin{figure}
\epsfig{file=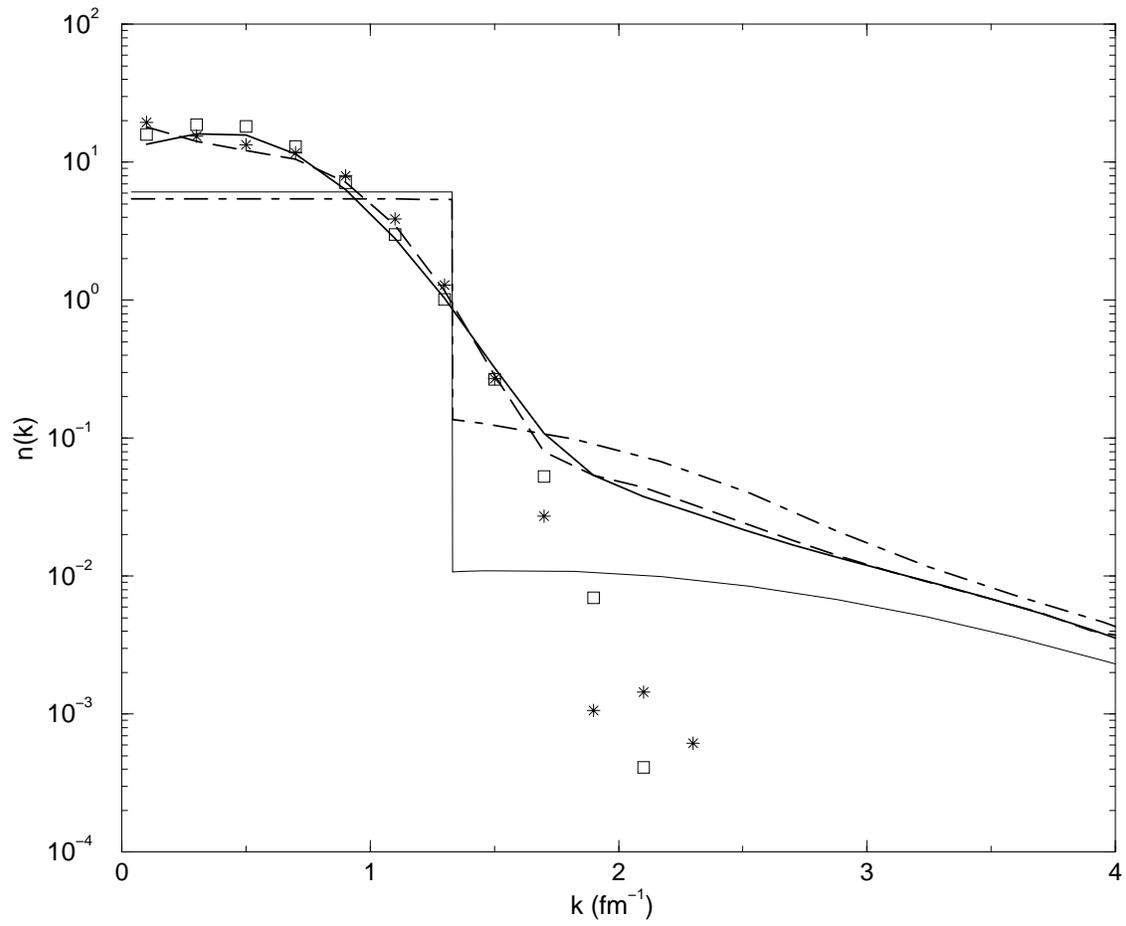,width=.75\textwidth,angle=-90}
    \caption{
 Correlated  momentum distributions in $^{16}$O, $^{40}$Ca and nuclear 
 matter (NM). See text.
      }
\label{fig:fig1} 
\end{figure}

The correlated momentum distributions computed within the CBF 
scheme are shown in Fig.\ref{fig:fig1}.  The MDs of $^{16}$O 
(continuous line) and of $^{40}$Ca (dashed line) are compared with
that of nuclear matter with Jastrow (thin solid line) and full 
(dash--dotted line) correlations. The IPM momentum distributions are given 
by squares ($^{16}$O) and stars ($^{40}$Ca). We stress that 
the differences between the Jastrow and the 
$f_6$ correlations are similar in the infinite and finite 
systems and that the three cases show an analogous behavior in 
large momentum region, that is dominated by the short range structure 
(and the NN correlations) of the nuclear wave function. 
The effect is, to a large extent, independent on the nucleus. 
The tensor correlations enhance the tails of the MDs by a factor 2--3 
with respect to the Jastrow case, slightly smaller than that 
found in Ref.\cite{pie92} ($\sim$4). 
The difference may be understood in terms of the stronger 
tensor force of the Argonne $v_{14}$ potential adopted in that reference, 
and of the presence of spin--orbit correlations, omitted in our calculation.

Another piece of information that can be extracted from the OBDM are the 
natural orbits  and their occupation numbers. The NO are obtained 
by diagonalizing the OBDM:
\begin{equation}
\rho_1({\bf r}_1,{\bf r}_{1'})=\sum_\alpha
 n_\alpha
 \phi_\alpha^{NO}({\bf r}_1)^\dagger
 \phi_\alpha^{NO}({\bf r}_{1'})
\, .
\label{natural}
\end{equation}

For spherical nuclei in ($ls$) single particle coupling, saturated
in both spin and isospin, this expression can be recast as:
\begin{equation}
\rho_1({\bf r}_1,{\bf r}_{1'})= \nu
\sum_l {{2l+1}\over {4\pi}}
P_l(\cos \theta_{11'}) \rho_l(r_1,r_{1'})
\, ,
\label{natural_1}
\end{equation}
where $P_l(x)$ are the Legendre polinomials, $\theta_{11'}$
is the angle between ${\bf r}_1$ and ${\bf r}_{1'}$ and 
$\nu$=4 is the nucleon degeneracy.  
Exploiting again the spherical symmetry, one obtains 
\begin{equation}
\rho_l(r_1,r_{1'})= \nu \sum_n n_{nl}\,
  \phi_{nl}^{NO}( r_1) \, \phi_{nl}^{NO}( r_{1'})
\, .
\label{natural_3}
\end{equation}

\begin{figure}
\epsfig{file=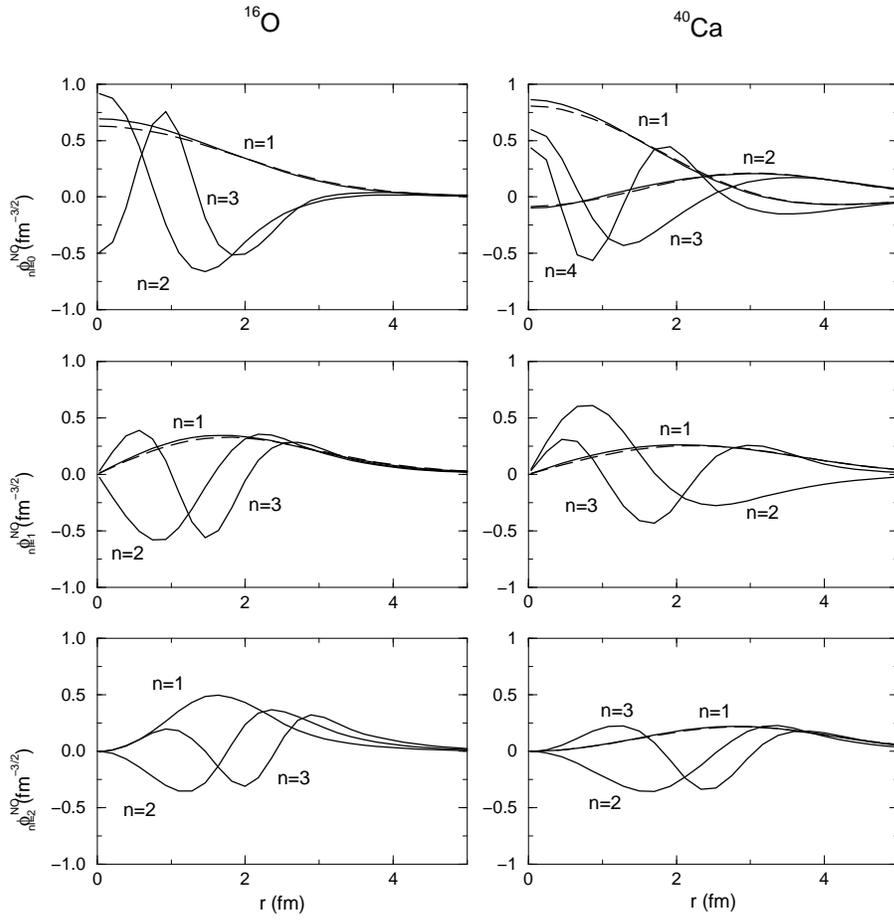,width=.75\textwidth,angle=-90}
    \caption{
Natural orbits of $^{16}$O and $^{40}$Ca.
Solid lines: $f_6$ model; dashed: IPM. 
      }
\label{fig:fig2} 
\end{figure}

The first three NO of $^{16}$O and $^{40}$Ca
 are shown in Figs. \ref{fig:fig2}. 
The NO occupation numbers  for Jastrow and $f_6$
correlations are given in Table \ref{tab:tab1} 
The effect of 
the correlations on the shell model orbitals are mainly visible in 
the short range part of the $1s$ state, making this NO more localized 
than its IPM counterpart. The shape of the other IPM states is 
barely influenced. The occupation of the 
NO corresponding to the shell model ones is depleted 
by as much as 22$\%$ (the $2s$ state in $^{40}$Ca). In contrast,  
the mean field unoccupied  states become sizeably populated. 
The two effects are largely due to the tensor correlations.  

The Green function (GF) approach of Ref.\cite{pol95} found 
the $^{16}$O $n=$1 NO more populated than the CBF ones for the 
occupied shell model states ($l=$0, 1), and, 
consequently, lower occupations for all the remaining orbitals 
($n_{1s}^{GF}$=.921,  $n_{1p}^{GF}$=.941 and  $n_{1d}^{GF}$=.017. 
The $1p$ ($1d$) occupation numbers are the 
average of the $1p_{1/2}$ and $1p_{3/2}$ 
($1d_{3/2}$ and $1d_{5/2}$) orbitals of the reference). 
These discrepancies are probably due to the different 
potentials adopted, rather than to the methodologies. In fact, 
in the GF calculation the one--boson--exchange Bonn B potential\cite{mac89} 
was used. The A8'+UIX model induces stronger correlations, so giving a 
larger depletion of the lowest NO. A similar effect was found 
in  $^3$He atomic drops\cite{lew88}, where 
the strong repulsive interaction between the $^3$He atoms 
depletes the shell model occupations by 15--46 $\%$.

\section{Single particle overlaps}

In a fixed center reference frame, the 
overlap  between the A--body ground--state and the (A-1)--body 
hole state of the residual system, $\psi_h$, is given by 
\begin{equation}
\psi_h(x) = {\sqrt A} {
{\langle \Psi_h(A-1)\vert \delta (x-x_A) \vert \Psi_0(A)\rangle}
\over
{\langle \Psi_h(A-1)\vert \Psi_h(A-1)\rangle ^{1/2}
 \langle \Psi_0(A)\vert \Psi_0(A)\rangle ^{1/2}}
}
\, .
\label{QH}
\end{equation}
In CBF theory $\Psi_h(1,2...A-1)$ is built by substituting 
$\Phi_0(1,2...A)$ in (\ref{Psi_0}) with  the Slater determinant 
obtained by removing a nucleon  in the state $h$,  $\Phi_h(1,2...A-1)$.
In doubly closed shell nuclei in the $ls$ coupling scheme
the radial dependence of the overlap function can be singled out, as 
\begin{equation}
\psi_h(x) = \psi_h(r) Y_{lm}({\hat r})\chi_{\sigma \tau}
\, ,
\label{psi_h}
\end{equation}
where $\chi_{\sigma \tau}$ is the spin--isospin single particle wave 
function.  In the IPM, the overlaps are simply the shell 
model functions and $\psi^{IPM}_h(r) = R_{h=nl}(r)$, where
$R_{h}(r_i)$ is the radial part of $\phi_\alpha(i)$. 

In order to develop a cluster expansion for $\psi_h(r)$, its 
expression is rearranged as:
 \begin{equation}
\psi_h(r) = {\cal X}_h(r) {\cal N}_h^{1/2}
\, ,
\label{QH1}
 \end{equation}
where
 \begin{equation}
 {\cal X}_h(r)= 
{\sqrt A} {
{\langle \Psi_h(A-1)\vert 
%{\cal Y}_{lm\sigma \tau}({\hat r}, {\bf \sigma}, {\bf \tau})
Y_{lm}({\hat r})\chi_{\sigma \tau}
\delta ({\bf r}-{\bf r}_A) \vert \Psi_0(A)\rangle }
\over
{\langle \Psi_h(A-1)\vert \Psi_h(A-1)\rangle }}
\, ,
\label{QH2}
 \end{equation}
and
 \begin{equation}
{\cal N}_h=
 {
{\langle \Psi_h(A-1)\vert \Psi_h(A-1)\rangle}
\over
{\langle \Psi_0(A)\vert \Psi_0(A)\rangle} }
 \, .
\label{QH3}
 \end{equation}
Cluster expansions and FHNC--like equations are then used to compute ${\cal X}_h$ 
and ${\cal N}_h$\cite{fab01}, along the lines given in Ref.\cite{ben89} to 
evaluate the overlap matrix elements in the CBF study of the nuclear matter 
spectral function.  

Finally, the spectroscopic factor, $S_h$, is given by the  
quasihole normalization, 
\begin{equation}
S_h=\int r^2 dr \psi_h^2(r)
\, .
\label{factor}
\end{equation}
The independent particle model gives $S_h^{IPM}=1$. 
Center of mass ($cm$) corrections are  a first source 
of correction. In fact, in the harmonic oscillator model they enhance $S_h$ 
for the valence hole states (those having the largest oscillator 
quantum number, $N_v$) by the $[A/(A-1)]^{N_v}$ factor\cite{die74}. 
So, the $cm$--corrected $1p$--shell spectroscopic factor of $^{16}$O is 
$S_{1p,cm}^{HO}=16/15\sim 1.07$, while the average between the $2s$ and $1d$ states 
in $^{40}$Ca is $S_{2s/1d,cm}^{HO}=(40/39)^2\sim 1.05$. Similar results 
are found with the more realistic Woods--Saxon orbitals\cite{nec98}.

The correlated spectroscopic factors (without $cm$ corrections) in 
$^{16}$O and $^{40}$Ca are shown in Table \ref{tab:tab2} for 
$f_6$ and Jastrow correlations. Jastrow correlations 
marginally reduce $S_h$ (at most 3$\%$).  Central spin--isospin 
correlations also provide a few percent depletion in 
the valence states, whereas the tensor correlations ($f_6$) give most of the 
reduction. For instance, they lower $S_{1p}$ in $^{16}$O to 0.90 and 
$S_{2s}$ and $S_{1d}$ in $^{40}$Ca to 0.86 and 0.87, respectively. 
The $1p$ CBF $^{16}$O result fully agrees with the variational Monte Carlo 
estimate of Ref.\cite{nec98}. The  operatorial correlations 
largely influence the low lying states, whose spectroscopic factors 
are drastically reduced by both central and tensor components:
$S_{1s}$ in $^{16}$O is 0.70, 
$S_{1p}$ and $S_{1s}$ in $^{40}$Ca are 0.58 and 0.55, respectively. 

The latest experimental values of $S_p$ from the 
$^{16}$O$(e,e'p)^{15}$N reaction\cite{leu94} are 
$S_{p_{1/2}}$=0.61 for the 1/2$^{-}$ ground state in $^{15}$N and  
$S_{p_{3/2}}(6.32)$=0.53 for the lowest 3/2$^{-}$  state at 6.32 MeV. This 
state exhibits 87$\%$ of the total $S_{p_{3/2}}$ strength, that is 
fragmented over three states at 6.32, 9.93 and 10.70 MeV. The total  
$S_{p_{3/2}}$ may be estimated to be $S_{p_{3/2}}$=0.53/0.87=0.61\cite{nec98}. 
In the $^{40}$Ca$(e,e'p)^{39}$K 
reaction\cite{lap93} the transition to the 1$d_{3/2}$ ground 
state gives  $S_{d_{3/2}}\sim$0.61$\pm$0.07. We recall that the 
CBF values are $S_{p}=$0.90 in $^{16}$O and  $S_{d}=$0.87 in $^{40}$Ca.

\begin{figure}
\epsfig{file=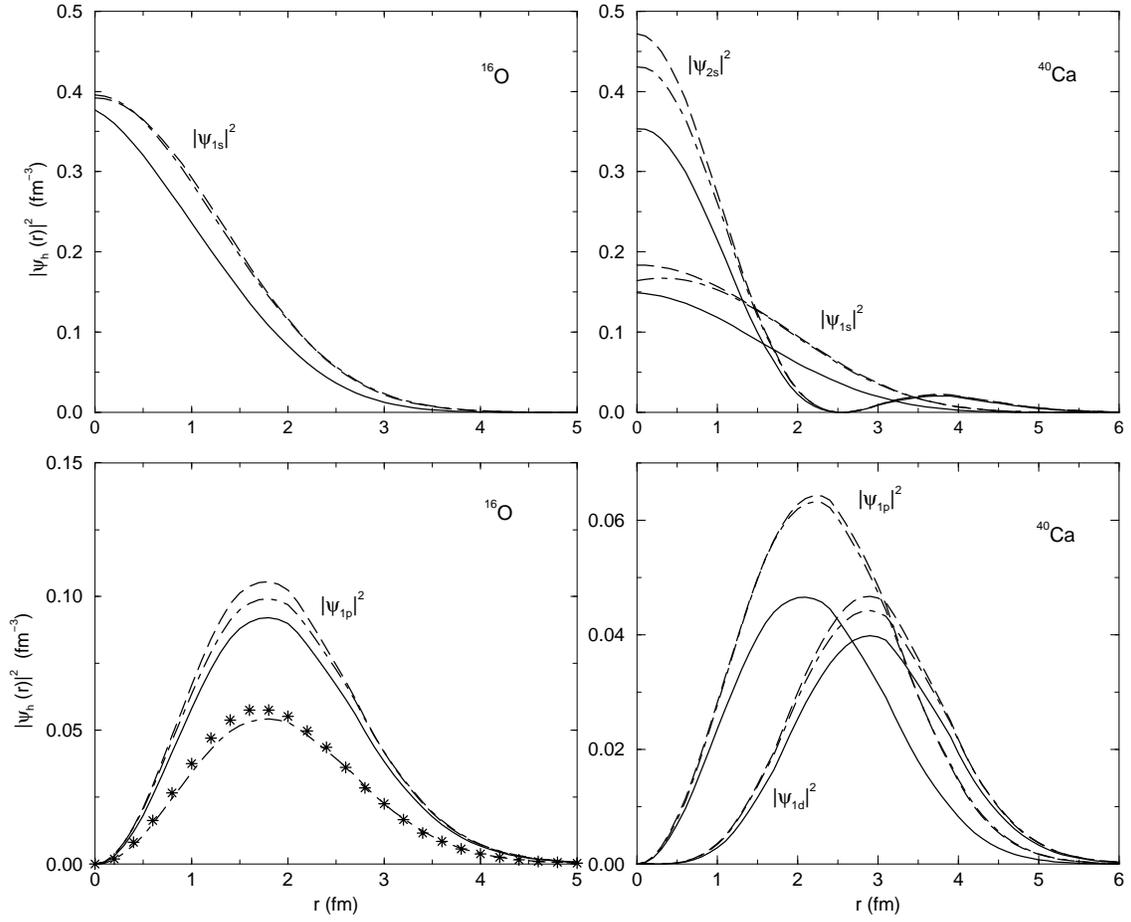,width=.75\textwidth,angle=-90}
    \caption{
Squared quasihole wave functions. 
Solid lines: $f_6$ model; dot--dashed: Jastrow; dashed: IPM. 
The $1p$ panel of $^{16}$O shows also the empirical overlap 
(stars) and the $f_6$ one, rescaled as explained 
in the text (lower dot--dashed line)
      }
\label{fig:fig3} 
\end{figure}

The squared overlap functions are shown in 
 Fig.\ref{fig:fig3} for the Jastrow and $f_6$ correlations, and are 
compared with the IPM ones. 
Jastrow components have little effect on the overlaps, 
while the spin--isospin correlations are the main responsible for the 
quenching of the IPM overlaps and of the spectroscopics factors.

 The IPM results obtained by a Woods--Saxon potential fitting  
the $^{16}$O$(e,e'p)^{15}$N cross section to the 6.32 MeV state with 
$S_{p_{3/2}}(6.32)$=0.53 are shown in the 
$\vert\psi_{1p}\vert^2$--$^{16}$O panel as stars. We have  
 rescaled $\vert\psi_{1p,FHNC}\vert^2$ by the factor 0.53/0.90 and 
the result is in nice agreement with the empirical estimate. 
\section{Conclusions}

In this contribution we have presented some results obtained within 
the correlated basis functions theory and concerning the properties 
of the single nucleon in the nuclear medium. More specifically, 
we have discussed the behavior of the one-body density matrix and 
some related quantities, as the momentum distributions and the 
 natural orbits, in  $^{16}$O and $^{40}$Ca using the Fermi 
hypernetted chain resummation technique. In addition, we have 
addressed the evaluation of the overlap functions and of the 
spectroscopic factors within the same approach. The relevance of the 
short range correlations, both of central and tensor type, has 
been stressed. The results presented here have been obtained  
using the Argonne $v_8'$ two--nucleon potential 
plus the Urbana IX three--nucleon interaction, together with a set of 
single particle wave functions fixed to reproduce at best the empirical charge
distributions of the two nuclei. 

The high momentum tail of the momentum distribution is largely 
dominated by correlations, and tensor components enhance the tail by 
a factor 3--4 with respect to the central ones. The tensor correlation is 
also important for the occupation of the natural orbits. In fact, the 
reduction of the occupation of the levels below the Fermi surface and, 
conversely, the enhancement above is amplified by the tensor terms. 
As far as the correlated overlap functions are concerned, they  are close 
to the single particle wave functions if central correlations are used, 
whereas their shapes are strongly modified by the tensor correlations.
The overall spectroscopic factors are depleted by 10-15\%, 
for the valence levels and 30-45\% for the deeply lying ones. 

In spite of this reduction, the FHNC approach in $^{16}$O 
does not reproduce the empirical $S_{p_{3/2}}$ spectroscopic 
factor extracted from $(e,e'p)$ reactions. A similar situation 
was met in Ref.\cite{ben90} for nuclear matter, where the variational 
calculation of the one--hole strength, $Z(e)$, around the Fermi level 
provided $Z_v(e\sim e_F)\sim$0.88,  mostly due to tensor correlations. 
Second order perturbative corrections in a correlated basis, including 
 two--hole one--particle, $(2h-1p)$,  correlated states, 
brought the strength to $Z_{CBF}(e\sim e_F)\sim 0.70$. This decrease 
explains almost half of the discrepancy with the empirical $^{208}$Pb 
spectroscopic factor, $Z(^{208}Pb)\sim 0.5-0.6$. 
The missing strength can be attributed to the coupling of
the single particle waves to the collective low-lying surface
vibrations, not reproducible in infinite nuclear matter.  
It is expected that  the inclusion of correlated 
$2h-1p$ corrections in the finite nuclei calculations 
can similarly take into account the coupling with surface vibrations.

\section*{Acknowledgments}
The results presented in this contribution have been obtained in 
collaboration with Giampaolo Co'. 
This work has been partially supported by MURST through the  
{\sl Progetto di Ricerca di Interesse Nazionale: 
Fisica teorica del nucleo atomico e dei sistemi a molticorpi}.

\newpage

\begin{table}

\caption{
Occupation numbers of the $nl$--th natural 
orbits for $^{16}$O and $^{40}$Ca in CBF, with the $f_6$ and 
Jastrow correlation models.
} 
\begin{tabular}{ccccc}
    $nl$ 
  &$n_{nl}$($f_6$;$^{16}$O) & $n_{nl}$(J;$^{16}$O)  
  &$n_{nl}$($f_6$;$^{40}$Ca)& $n_{nl}$(J;$^{40}$Ca) 
 \\ \\
\tableline
 \\ 
 $1s$ & 0.858 & 0.960 & 0.864 & 0.952 \\
 $2s$ & 0.019 & 0.005 & 0.780 & 0.962 \\
 $3s$ & 0.010 & 0.002 & 0.052 & 0.002 \\
 $4s$ & 0.005 & 0.001 & 0.013 & 0.001 \\
      &       &       &       &       \\
 $1p$ & 0.919 & 0.980 & 0.841 & 0.949 \\
 $2p$ & 0.021 & 0.004 & 0.024 & 0.009 \\
 $3p$ & 0.011 & 0.003 & 0.016 & 0.006 \\
      &       &       &       &       \\
 $1d$ & 0.025 & 0.006 & 0.956 & 0.983 \\
 $2d$ & 0.011 & 0.003 & 0.030 & 0.007 \\
 $3d$ & 0.006 & 0.001 & 0.019 & 0.006 \\
 \\ 
\tableline
\end{tabular}
\label{tab:tab1}
\end{table}

\begin{table}
\caption{
CBF spectroscopic factors for $^{16}$O and $^{40}$Ca, with Jastrow (J) and 
$f_6$ correlations.
} 
\begin{tabular}{cccccccccc}
  & corr. &$1s$ &$1p$ &$1d$ &$2s$ 
 \\
 \\
\tableline
 \\
 $^{16}$O &   J   & 0.98 & 0.98 &      &      
 \\
          & $f_6$ & 0.70 & 0.90 &      &      
 \\
 \\
 $^{40}$Ca&   J   & 0.98 & 0.99 & 0.97 & 0.98 
 \\
          & $f_6$ & 0.55 & 0.58 & 0.87 & 0.86 
 \\
 \\
\tableline
\end{tabular}
\label{tab:tab2}
\end{table}

\newpage

\end{document}